\documentclass[11pt]{article}

\usepackage{latexsym,amsmath,amssymb,theorem,epsfig}
\def\url#1{\expandafter\string\csname #1\endcsname}

\numberwithin{equation}{section}

\def\a{\alpha}
\def\b{\beta}
\def\g{\gamma}

\def\d{\delta}

\def\f{\phi}

\def\k{\kappa}
\def\l{\lambda}
\def\m{\mu}

\def\p{\pi}

\def\s{\sigma}
\def\t{\tau}

\def\G{\Gamma}

\def\L{\Lambda}

\def\P{\Pi}

\def\be{\begin{equation}}
\def\ee{\end{equation}}
\def\bea{\begin{eqnarray}}
\def\eea{\end{eqnarray}}
\def \ba {\begin{align}}
\def \ea  {\end {align} }

\def \td {\tilde}

\def \ha {\tfrac{1}{ 2}}

\def \del{\partial}
\def \a {\alpha}

\def\ov{\over}
\def \ci {\cite}

\def \foot {\footnote}
\def \bi{\bibitem}
\def\la{\label}\def\foot{\footnote}\newcommand{\rf}[1]{(\ref{#1})}

 \def \bg {\bar g}
\def \OO {{\cal  O}}\def \no {\nonumber}

\def \LL {{\rm L}}
\def \N {{\cal N}}

\def \g {{\gamma} } 
\def \tdp {\td \p}

\def \vp {\varphi}  \def \te {\textstyle}

\def \G {\Gamma} \def \ba {\begin{align}}

\def \edo {\end{document}}

\def \N {{\cal N}}
\def \VV   {{\rm V}} 
\def \Tr {{\rm Tr}}
\begin{document}
\def \p {\phi}

\def \four{\tfrac{1}{4}}

\def \vpn { \vp_{\rm in}}
\def \tf {\tfrac}  \def \LL  {{\cal L}}  \def \VV {{\cal V}}
\def \s {\sigma} \def \t {\tau} 
\def \G {\Gamma} 
\def \p {\phi} 

\def \f  {\frac}\def \cD {{\cal D}}\def \ne {\nabla} \def \Im {{\rm Im\,}}
\def \eps {\epsilon}
\def \iffa  {\iffalse}
\def \d {\delta} 
\def \pn  {\p_{\rm in}}\def \P {\Phi}

 \def \A {{\cal A}}\def \k {\kappa} 
\def \bi {\bibitem}
\def \PP {{\rm P}}

\textwidth 170mm 
\textheight 230mm 
\topmargin -1cm
\oddsidemargin-0.8cm \evensidemargin -0.8cm 
\topskip 9mm 
\headsep9pt

\overfullrule=0pt
\parskip=2pt
\parindent=12pt
\headheight=0in \headsep=0in \topmargin=0in \oddsidemargin=0in

\vspace{ -3cm} \thispagestyle{empty} \vspace{-1cm}
\begin{flushright} 
\end{flushright}
 \vspace{-1cm}
\begin{flushright} Imperial-TP-AT-2022-06\\   \date {  } \today 
\end{flushright}
\begin{center}

 \vspace{1.2cm}

{\Large\bf
Comments on  4-derivative scalar theory in 4 dimensions
 }
 
 \vspace{0.8cm} {
  A.A.~Tseytlin\footnote{Also  on leave  from  the Institute for Theoretical and Mathematical Physics (ITMP) and Lebedev Institute.
    }  
   }\\
 \vskip  0.5cm

\small
{\em
  Theoretical Physics Group, Blackett Laboratory\\ Imperial College London,
 SW7 2AZ, U.K. }

\normalsize
\end{center}

 \vskip 1.2cm

 \begin{abstract}
 We review and elaborate on some aspects of the classically scale-invariant renormalizable 4-derivative  scalar theory  $L= \phi\,  \partial^4 \phi + g (\partial \phi)^4$.  Similar models  appear, e.g.,  in the context  of conformal supergravity or in the description 
 of  crystalline  phase of membranes. Considering this theory in Minkowski signature we suggest  how to  define Poincare-invariant   scattering amplitudes by assuming that only massless  oscillating  (non-growing) modes  appear  as external states. In such shift-symmetric  interacting theory  there are no IR divergences despite  the presence of   $1/q^4$   internal propagators. We discuss how non-unitarity of this theory manifests itself  at the level of the one-loop  massless scattering amplitude.

 \iffa 

Comments on  4-derivative scalar theory in 4 dimensions

A.A.~Tseytlin

Comments:    19 p.,  based on talk at A. Slavnov  memorial conference  (21-22.12.2022)

\fi

\end{abstract}
\vskip 0.8cm

\newpage

\tableofcontents

\renewcommand{\theequation}{1.\arabic{equation}}
 \setcounter{equation}{0}
\setcounter{footnote}{0}
\section{Introduction}

Despite its apparent non-unitarity  the $\p\Box^2\p$ theory  for a dimensionless  scalar  $\p$  in 4 dimensions  attracted attention 
in the past   \ci{Narnhofer:1978sw,Binegar:1983kb,dEmilio:1982ghe}) and also recently 
(see, e.g., \ci{boxes,Brust:2016gjy,Adamo:2018srx,Gibbons:2019lmj,Romoli:2021hre,Boyle:2021jaz,Safari:2021ocb,
Stergiou:2022qqj,Chalabi:2022qit,Buccio:2022egr}).

Focussing on  the  free theory  may   not be particularly illuminating  as  an interpretation  of the  theory may depend on 
allowed interactions  and   types of observables  considered.\foot{One may wonder if   problems  of 
the free   theory (related to higher time derivatives and   non-positivity of the energy)  may 
 be cured   by a special choice of initial conditions  or   assuming 
 propagation in a non-trivial  curved background  \ci{Gibbons:2019lmj}  
but  the main  issue  is  what happens  at the interaction level (cf.  \ci{Smilga:2017arl}).  }
Here we   will focus on the following classically scale invariant 
renormalizable theory of  a real dimension 0  scalar with 
the Euclidean action 
\be 
S= \int d^4 x\,    L_4 \ , \qquad \ \ \  L_4=    (\del^2 \p)^2 + g (\del^m \p\, \del_m \p)^2   \ . \la{01} \ee
Here $g$ is dimensionless   coupling  (we assume $g >0$     for the Euclidean action to be positive).\foot{ 
We will  be assuming that  a (quadratically divergent)  coefficient of the 2-derivative term $(\del \p)^2$  that is ``induced"  if the model \rf{01} 
is regularized  using a dimensionfull   cutoff is  fine-tuned to 0  after the renormalization.
 Below we will  use  dimensional regularization in which  power divergences do not appear and thus \rf{01}  is renormalizable   without 2-derivative term.}
While this  may be 
 ``unnatural",   similar  fine tuned 
 (low-dimensional) models appear  in the  description of  the  crystalline  phase of membranes 
(see, e.g.,  \cite{David:1987si,Bowick:1999rk,Coquand:2020tgb}).\foot{\la{f1}The free energy   of a  membrane  is 
$\int  d^d x \Big[ T ( \del_m  X^a) ^2  + \kappa  (\del^2 X^a )^2 +  \l ( \del_m X^a \del_m  X^a)^2 + \mu  ( \del_m X^a \del_n X^a)^2 \Big] $ 
where $X^a$ ($a=1, ..., N$) is an  embedding coordinate ($d=2$  and $N=3$  for a standard membrane). 
$T$ is the tension, $\l, \mu $ are the   elastic   constants (Lame coefficients)  and $\kappa$  is the 
 bending rigidity (the coefficient of extrinsic curvature coupling). After the crumpling transition   from  the  elastic phase 
to the   crystalline  membrane   phase one has $T\to 0$. 
Setting $X^a= (x^n  + u^n(x),  h^\a(x))  $ and    integrating over $u^n$ one 
gets an effective action for the transverse coordinates $h^\a$.   In 
 the formal limit of  $N=1, \ d=4 $  the  above  energy  functional  is  the same  as the Euclidean  action for  \rf{01}.}

The  4-derivative  scalar  model  \rf{01}   appears   also in the context of 
  extended conformal supergravity \cite{Bergshoeff:1980is,Fradkin:1981jc,Fradkin:1985am,Butter:2013lta} as  a natural partner of  the Weyl graviton (see also \ci{Berkovits:2004jj,Beccaria:2016syk,Adamo:2018srx}).  
 It also appears as a part of the effective action for  the integrated 4d conformal   anomaly   \ci{Fradkin:1983tg,Komargodski:2011vj}  (with $\p$  interpreted as  a  conformal factor of the 4d metric).
 Related to the  fact that  the conformal supergravity  may be interpreted as 
   (the logarithmically divergent  part of) an induced action 
 from $\N=4$  SYM theory, one can get a similar scalar  action 
  by  considering  the  Maxwell theory in the  background of the 
 complex local  coupling  $\tau = C + i e^{-\phi}$  and curved 
 metric. Starting with $ e^{-\p} F_{mn} F^{mn} + i C F^*_{mn} F^{mn}        $ 
 and integrating out the vector field 
 one finds that the  resulting  log  UV divergence  is proportional to the $SL(2,R)$ covariant   Lagrangian \cite{Osborn:2003vk,Buchbinder:2012uh}
\ba
& L=\tfrac{1}{4(\Im\tau)^2}\Big[\cD^2\tau
\cD^2\bar\tau-2(R_{mn}-   \tfrac13 R g_{mn})  \ne^m\tau\ne^n\bar\tau \Big]
\cr
&
\ \ \ \ \ \ \ \ \  +\tfrac{1}{48(\Im\tau)^4}\Big( \ne^m\tau\ne_m\tau\ne^n\bar\tau\ne_n\bar\tau+
2    \ne^m\tau\ne_m\bar\tau   \ne^n\tau\ne_n\bar\tau    \Big)  \ ,  \la{03} \\
& \ \ \ \cD^2\tau\equiv \nabla^2 \tau+\tfrac{i}{\Im\tau}\ne^m\tau\ne_m\tau
,\ \ \ \
  \cD^2\bar\tau\equiv \nabla^2 \bar\tau-
\tfrac{i}{\Im\tau}\ne^m\bar\tau\ne_m\bar\tau \ . \no \end{align}
For  $C=0$ or $\tau=  i e^{-\phi}$   this reduces in flat space  to the real scalar  model  like \rf{01}.

 To address the question of how to define  observables in the theory   \rf{01}
 we will   consider the analog of  massless  on-shell    scattering amplitudes  with the aim 
  to see  how  the expected  non-unitarity  of the model \rf{01}  is reflected in the S-matrix. 
 The subtlety of the ``dipole ghost"  theory  with  $\p \Box^2\p $   kinetic term  ($\Box\equiv \del^m\del_m= - \del_0^2 + \del_i^2$)
  is that it  is not a smooth limit 
 of the ``massive ghost"    $\p ( \Box^2 + \mu \Box) \p $ 
  model   which 
 may be viewed as describing  a ``diagonal" combination of  a standard  2-derivative 
 massless scalar  and  a ghost-like massive  scalar    and which is thus obviously non-unitary
 (unless one resorts to some special prescriptions, cf. \ci{rr}). 
 Indeed,   in introducing an auxiliary field $\psi$    so  that 
  $  \p  \Box^2  \p \to  2 \psi \Box \p - \psi^2$
 one  may define an equivalent  model 
 $
  2 \psi \Box \p - \psi^2  + \mu \p \Box \p$   which can be  diagonalised in terms of   
 $\vp = \p + \mu^{-1} \psi$ and 
  $\psi$  as   $ \mu \vp \Box \vp - \mu^{-1} \psi ( \Box + \mu ) \psi$. This  diagonalization  
  becomes  singular in the limit $\mu\to 0$  which is a manifestation of the fact   that  $\p \Box^2 \p$   or 
    $ 2 \psi \Box \p - \psi^2$ describes a ``non-decomposable"   system (cf. \cite{Binegar:1983kb}).
    
  It  is interesting to note that this  non-diagonalizability  is lifted if one starts with  the Weyl-invariant analog 
  of the $\Box^2$  operator in curved space (cf. \rf{03})    \ci{Fradkin:1981jc,Fradkin:1985am}:  $ (\Box \p)^2 \to 
  \nabla^2 \p \, \nabla^2 \p - 2 (R_{mn } -  \tfrac13 R g_{mn})  \ne^m\p \ne^n\p $. 
  In the case of  an  Einstein  space background $R_{mn} =  \tfrac14 R g_{mn}$ (e.g. 4-sphere or AdS$_4$)
  this  reduces   to $\p  ( \ne^2  - \tfrac16 R ) \ne^2  \p$  which can be diagonalized as above in to a combination of a physical massless $\ne^2$ scalar and a  ghost-like  conformal  $-\ne^2 + \tfrac16 R$ scalar    at the expense of  introducing 
  $R^{-1}$  factors that are singular in the flat space limit. 
  
 Returning to the flat space case, 
 the  family of solutions  of $\Box^2 \p=0$  contains in addition to  ``massless" oscillating solutions  of $\Box \p=0$ 
 (or  $\p(x)  \sim  \td \p(p) e^{ip\cdot x}, \ p^2=0$)  also  ``growing'' solutions $\p(x) \sim  N_n(p)\,  x^n e^{i p\cdot x}$ 
 (cf.  also  \ci{Berkovits:2004jj}).
 While the space  of the corresponding states can not be diagonalized\foot{The free ``dipole ghost"  theory  corresponds to a non-unitary representation of the conformal group $SO(2,4)$ 
 \cite{Binegar:1983kb}, 
  both in the context of a   Euclidean  theory and  Minkowski   theory; 
  the corresponding states   belong to a non-diagonalizable  module.}
 it   appears to be  consistent to  define the  scattering amplitudes   with only the oscillating modes appearing 
  on the external lines.  
 The analogs of the  scattering   amplitudes  for the growing modes  are  not well defined  \cite{Adamo:2018srx}:  they are  IR divergent 
 and not conserving momentum (resembling scattering  in an external field). 
 
 We shall thus  focus on  the scattering amplitudes for a subclass $\Box \p=0$  of the solutions  of $\Box^2 \p=0$ 
 as defining the asymptotic states.  
 Since 
   the internal propagators    $\Box^{-2} \sim \log x^2$   do  not decay at large distances,  
 the question of  whether  the resulting  amplitudes   are well defined  (finite in the IR)  crucially depends 
 on type of interactions one is going  to considers. This is  analogous  to  what happens  in  the  massless 
 theory in 2 dimensions where   also $\Box^{-1} \sim \log x^2$,   leading   to  the 
 familiar   dogma  that  ``massless S-matrix  does not exist in 2 dimensions". 
 This actually   applies  only to  local non-derivative interactions  while 
 the S-matrix  in a theory  like $\p \Box \p + V(\del \p)$   (which is invariant
 under the shift symmetry 
 $\p \to \p + c$  and thus  having only $\langle \del_m \p \, \del_n \p \rangle$ correlators that decay at large distances)\foot{An example    \ci{Dubovsky:2012wk} 
 is  the  Nambu  action in the static   gauge   (in Euclidean signature) \\
$
L= \sqrt {  \det ( \delta_{ij} +  \del_i \p^a \del_j \p^a ) } = 1 + \ha   \del^i \p^a \del_i\p^a   + {1\ov 8} 
\big[ (\del^i \p^a \del_i \p^a )^2  -    2   ( \del_i \p^a \del_j \p^a)^2\big] +  O((\del \p)^6)
$.}   is  well defined  in the IR. 
 The same  will  apply  to the $\Box^2$  theory \rf{01} in 4 dimensions.

 It  will be  instructive to compare the  renormalizable model \rf{01} with a  similar 4d model with the  standard kinetic term 
 \be 
   L_2=   \p \del^2 \p + \bg (\del^m \p\, \del_m \p)^2   \ , \ \ \ \ \ \  \ \ \ \ \  \bar g = M^{-4}   g \ ,  \la{02} \ee
which requires higher order counterterms   but   may be  treated as an effective  field theory (see, e.g.,   \cite{Weinberg:1995mt,Adams:2006sv}). Here 4d field $\p$ has dimension 1  and    $M$ is a   mass scale ($g$ is dimensionless). 
This theory is unitary in the   low-energy perturbation theory  (i.e. assuming $s=E^2 <  M^2$). 
This  implies, in particular,  the  validity  of the generalized  optical theorem: 
the imaginary part  of the one-loop  4-particle scattering amplitude will be related to  (the  phase-space integral of) the square 
of the tree level amplitude  given by 
$A_4^{(\rm tree)} \sim \bg (s^2 + t^2 + u^2)$. 

The massless 
 tree  level  amplitude in the theory \rf{01}   (constructed  using   the standard rules  \cite{Arefeva:1974jv},   i.e. 
  by evaluating the classical action on the solution of the equations of motion  for \rf{01}   
  with $\p=\p_{\rm in} + \OO(g), \ \Box \p_{\rm in} =0$) 
 will be given   by the same  expression  as in the model  \rf{02} with $\bg \to g$, i.e.  $A_4^{ (\rm tree)} \sim g (s^2 + t^2 + u^2)$. 
 The expression for the 
 one-loop  amplitude, however,  will  involve not the standard $1\ov q^2$   but  the $1\ov q^4$   internal propagators. 
 As we shall see below, its imaginary part will   be proportional to the first  rather than second power of  the tree   amplitude, 
in conflict with  the generalized   optical theorem.
 This is, of course, hardly surprising given that 
 restricting the external states  to the massless only  these are not the   ones 
  propagating on internal lines:
the  $1\ov q^4$ propagator effectively describes an irreducible  mixture of  the massless   and ``growing" modes. 

This  may  be contrasted to the massive ghost theory  $\p( \Box^2 +\mu  \Box) \p + ...$ 
with the propagator $  {1\ov q^4 -\m q^2} = {1\ov \mu } ( {1\ov q^2} - {1\ov q^2 -\m} ) $. 
Considering   here    the 
 amplitudes for   the massless  modes only  we will   have both massless   and massive ghost 
states propagating on the internal  lines  so cutting the 
 internal lines   will  relate the imaginary part  of the one-loop amplitude 
to tree level amplitudes of  both massless and massive states  (with the 
 breakdown of unitarity related to negative norms of the ghost states).

 In the  ``dipole ghost" model  where only  the massless   scattering amplitudes are  well defined 
 (non-singular in the IR) we  get a conflict 
 with   the optical theorem   which is  effectively   due to  inability to implement the
  completeness relation  on the space of states.  
  
One  may still wonder  if  there is a   possible  generalization of the  notion of unitarity that may still apply 
in this case.  
For example, one could  try to modify (i)  the prescription  of summing over    phase space of intermediate states
or (ii)  the Minkowski   continuation of the one-loop diagram  or $i \eps$ prescription for computing 
its  imaginary part. 

We will not be able to address  this  question here,    limiting  our   goal
 to just  computing  explicitly the one-loop  massless 
scattering amplitude in the theory \rf{01}  and  discussing  
 the violation of the standard   version of the optical  theorem.

As another indication of the difference between the   theories \rf{01} and   \rf{02} 
let us  recall the argument  \ci{Adams:2006sv} (see also \cite{CarrilloGonzalez:2022fwg})
relating the  positivity  of  the coupling $\bg$ in \rf{02} to   the condition of causal (subluminal) 
propagation of  small perturbations in  the  classical background $\p_0= u_m x^m, \ u_m$=const. 
Expanding $\p = \p_0 + \td \p$  we find that $L_2$  in \rf{02}  takes the form (in Minkowski signature) 
\be \la{04}
L_2 = - K^{mn} \del_m \td \p\,  \del_n  \td \p + \OO((\del \td\p)^3) \ , \ \ \ \  \ \ \ \ \ 
 K_{mn} = (1 - 2\bg u^2)\eta_{mn}   -   4 \bg u_m u_n \ . 
\ee
The corresponding dispersion relation in momentum space  ($\del_m \to i p_m$) is 
\be   (1- 2\bg u^2) p^2 -  4 \bg (u^m p_n)^2 =0 \ . \la{05} \ee
Assuming that  in perturbation theory $1 - 2\bg u^2 >0$  one  concludes that  to have  subluminal propagation
(i.e. 
$ p_0^2 = v^2 {\vec p}\,^2 $  with $  v^2 < 1$  or  $p^2 \equiv  -   p_0^2 +  \vec p\,^2=  (1 -v^2)  \vec p\,^2 \geq  0$) 
 one  should have $\bg >0$.  
The  same argument  repeated  for $L_4$  in \rf{01} 
gives instead of \rf{04},\rf{05} 
\be \la{06} 
 K_{mn} = \eta_{mn} (\Box - 2g u^2)  -   4 g u_m u_n \ , \ \ \ \ \ \ \ \ 
 p^2 (p^2+ 2 g u^2) +  4 g (u^m p_n)^2 =0 \ .  \ee
Here for $g >0$  (required for positivity of the Euclidean action in \rf{01})
the  subluminal $p^2 >0$   solution may   exist  only if  $u_m$ is time-like ($u^2 <0$). 
This suggests  breakdown of causality (and related analyticity properties of  S-matrix) in the  theory \rf{01}. 


Below   in section 2   we shall first  define  the one-loop effective action 
 and compute  the corresponding beta-function in the theory \rf{01} and  its  multi-scalar  generalization. 
 In section 3.1   we shall  motivate the definition of the massless S-matrix 
  starting   with the 2-derivative action   equivalent to \rf{01}  and then
    in section 3.2  compute the  explicit expression  for the one-loop   scattering amplitude.
  In section  3.3 we shall  review  the generalized  optical   theorem  explaining 
   why it is valid in the standard $\p^4$ theory and also  in the  unitary 
   model \rf{02}  but fails in its  renormalizable  analog \rf{01}. Some concluding remarks will be made in section 4.

\renewcommand{\theequation}{2.\arabic{equation}}
 \setcounter{equation}{0}


\section{One-loop effective action  and     beta-function}
 
 Starting with the action \rf{01} 
one can  compute the one-loop effective action   for a  generic background.  One can then  
determine  the  renormalization of the coupling $g$ and also find the  one-loop  scattering amplitudes.  
 Setting $\p= \vp + \tdp$  where $\vp$ is a classical  background  and expanding to  quadratic order in $\tdp$ we get (here we consider the Euclidean signature) 
\begin{align}
& L_4 = \del^2 \tdp\,  \del^2 \tdp  -   V^{mn} (\vp) \del_m \tdp \del_n  \tdp+ O(\tdp^3)\ ,  \la{2} \\
& V_{mn} (\vp)  =- 2 g (   \delta_{mn} \del^k \vp \del_k \vp   + 2 \del_m \vp \del_n  \vp) \ . \la{3}
\end{align}
Then the   one-loop effective action is 
\be \la{44}
\Gamma_1 = \ha \log \det \Big[ \del^4 + \del_m  \big(V_{mn}(\vp)  \del_n  \big) \Big]  \ . \ee
Ignoring quadratic divergence proportional to $ V_m^m$  the logarithmic one is given  by 
\cite{Fradkin:1981iu} ($\L\to \infty$)
\be \la{33} 
(\Gamma_1)_\infty = -  \tfrac{1}{(4 \pi)^2}  \log \tfrac{ \Lambda}{ \mu}  \int d^4 x \,  b_4 \ , 
\qquad \ \   b_4= \te   {1\ov 24 } V_{mn} V_{mn}   + {1\ov 48}  (V^m_m)^2 \ .  
\ee 
From \rf{3}   we get 
\be \la{4}
b_4 =  5 g^2 (\del_m \vp \del_m \vp)^2 \ . 
\ee
This divergence can be absorbed   into the  renormalization of the   coupling: 
$g_b(\Lambda) -   {1\ov (4 \pi)^2}   5  g_b ^2 \log { \Lambda\ov \mu}  = g(\mu)$. The 
 resulting RG equation  for the renormalized coupling  is\foot{The same beta-function  was found in \ci{Bowick:1999rk}  and in \cite{Safari:2021ocb}.}
\be
{d g\ov d t } =\te  {1\ov (4 \pi)^2}  5 g^2 \ , \ \ \ \  \qquad  t= \log \mu\ .    \la{5} \ee
Thus $g\to 0 $  in the IR  ($\mu  \to 0$) and grows in the UV. Thus   the theory \rf{01}  with  $g >0$ 
 (i.e. with  positive Euclidean action) 
is   similar to the standard $\p^4$ model  in not  being 
   asymptotically free  and thus  not defined  at short   scales beyond the Landau pole. 


It is straightforward to generalize  this discussion to   the  analog of the model  \rf{01} 
with several   scalar fields $\p^a$, \ \  $a=1,..., N$.  For $N >1$ 
 there are two   independent  quartic invariants, i.e.   \rf{01} is generalized to (cf. footnote \ref{f1}) 
\be \la{6}
 L=\del^2 \p^a \del^2 \p^a +   g_1  (\del^n \p^a \del_n \p^a)^2  + g_2   ( \del^n  \p^a \del^m \p^a) ( \del_ n \p^b \del_m \p^b)   \ . 
 \ee
 For $N=1$ this  reduces  to \rf{01}  with $g= g_1 + g_2$. 
Setting   $\p_a= \vp_a + \tdp_a$  as in \rf{2} gives  for the quadratic action (which generalizes \rf{2},\rf{3}) 
\ba  
& \qquad \qquad  L= \del^2 \tdp_a \del^2 \tdp_a  -   V^{ab} _{mn} (\vp) \del_m \tdp_a \del_n  \tdp_b  + O(\tdp^3) \ , \la{2a} \\
& V^{ab}_{mn} = 
- 2 g_1 ( \delta_{ab}   \delta_{mn} \del^k \vp_c \del_k \vp_c    +  2 \del_m \vp_{(a}  \del_n \vp_{b)}    )
   - 2 g_2     ( \delta_{ab}    \del_m \vp_c \del_n \vp_c
      +   \delta_{mn}  \del^k \vp_a  \del_k \vp_b     +     \del_m \vp_{(b}  \del_n \vp_{a)}     ) \ . 
\no  
\end{align} 
In particular, 
$
 (V^{ab} )^m_m  = - 2 g_1 ( 4\delta_{ab}   \del^m \vp_c \del_m \vp_c    + 2 \del^m \vp_a  \del_m \vp_b      )
   - 2 g_2     ( \delta_{ab}    \del^m \vp_c \del_m \vp_c      + 5 \del^m \vp_a  \del_m \vp_b      ).$
The  coefficient of the logarithmic divergence in \rf{33}  is, in general, 
\be \la{311}
b_4= \te  {1\ov 24 } V^{ab}_{mn} V^{ab}_{mn}   + {1\ov 48}  (V^{ab})^m_m  (V^{ab} )^n_n \ , 
\ee 
so   that the  resulting beta-functions for $g_1$ and $g_2$  are found to be\foot{These are in  agreement 
with   expressions    in   with \ci{Bowick:1999rk} 
after setting  there $d=N, \ D=4$  and  $g_1=  4v - u, \ g_2 = 4 u$   where $u$ and $v$ are 
the two  couplings in \ci{Bowick:1999rk} (accounting also  for the 1/2  normalization of  the quadratic term in \ci{Bowick:1999rk}).
Higher-loop renormalization of similar  membrane models  was discussed in 
\cite{Coquand:2020tgb}.}  
\begin{align}
& \la{312} {d g_1\ov d t } = \te {1\ov  (4 \pi)^2} \Big[ 2(N + {7 \ov 6}     )  g_1 ^2    +  (  N   + {17\ov 3})  g_1  g_2   +  { 1 \ov 12} (N  +15)g_2^2   \Big]   \ ,    \\
&\la{313}  {d g_2\ov d t } = \te {1\ov   (4 \pi)^2}   \Big[  {2 \ov 3}   g_1 ^2    + {10 \ov 3}  g_1  g_2   +   { 1 \ov 6} (N + 21) g_2^2   \Big]\ . 
\end{align} 
For   $N=1$   these  equations imply \rf{5} 
 for $g=g_1 +g_2 $.
 
  The beta-functions in \rf{312},\rf{313}   have no   zeroes 
  if    $g_1 $ and $g_2$  are positive 
   and, in general,  no   common zeroes   so that 
    both $g_1$ and $g_2$    are not asymptotically free, i.e.  run to a Landau pole  singularity   in the UV. 
        It may be of interest to  consider a supersymmetric   generalization of the  model \rf{6} (cf. \cite{Butter:2013lta})
     to see  if in this case the corresponding RG equations may have  non-trivial fixed-point solutions. 
    


\renewcommand{\theequation}{3.\arabic{equation}}
 \setcounter{equation}{0}

\section{Scattering amplitudes}

Let us now   consider the theory \rf{01}   in Minkowski space and address the issue  of 
how to compute  the corresponding 
scattering   amplitudes.  
One possible strategy is to compute   the effective action and then 
evaluate it on a classical solution with appropriate   asymptotic  (or $g \to 0$)  behaviour.

As was mentioned in the Introduction, while   the $\Box^2\p=0$ equation admits  both oscillating  and growing solutions,  only the  former can be used 
 as asymptotic  ones  as  otherwise  the analogs of the scattering amplitudes will   be  IR divergent 
\ci{Adamo:2018srx}. 
The restriction   to oscillating modes as external states 
 follows naturally from the ``2-derivative" formulation of the  model \rf{01} that we shall discuss first. 

\iffa 
Instead of trying to define the space of states of free $\Box^2$ theory one may  follow formal approach:  
 compute one-loop effective  action  for any background;  evaluate it on a 
  classical solution  of $\Box^2 \p+ ...=0$; expand  to given order in external legs, etc. 
The fact   that interaction term is   invariant under constant shift of $\p$ is   necessary 
 for  getting   an IR   finite S-matrix  despite  the fact that the propagator $\sim log x^2$ is  growing in the IR. 
This is similar to what happens   in  2  dimensions  in  Nambu-like  action  \ci{Dubovsky:2012wk}. 

In contrast to 2d  case (where the free equation is $\Box\p=0$ and thus have only standard wave-like solutions $\p \sim \td \p(p)  e^{ipx}, \ p^2=0$)  here there are also   growing modes $\p\sim N_n(p) x^n e^{ipx}$
(see, e.g., \ci{Berkovits:2004jj,Adamo:2018srx})  for which  there is no natural notion  of S-matrix. For example, 
 the  analogs of tree-level scattering amplitudes   extracted from the classical action evaluated on  the corresponding   solution do not respect momentum conservation  and  are divergent in the IR  \ci{Adamo:2018srx}. 
 \fi 
 
 
\subsection{2-derivative formulation} 


Let us  start with   the Lagrangian 
\be \la{222}
L(\p,\psi)=   2  \psi \del^2  \p   - \psi^2    + g  (  \del^m   \p\, \del_m \p)^2  \ ,  
\ee
from which  the $\Box^2$ model \rf{01} follows  upon   integrating out the  dimension 2  field $\psi$. 
The two   theories \rf{01} and \rf{222}
 are   equivalent  as long as we do not   introduce sources   for  $\psi$, i.e. consider only
observables   built out of  the  dimension 0 scalar  $\p$. 
The  S-matrix  found  from  \rf{222} and restricted to the $\p$-sector only   should thus be the same as  the one found from 
  \rf{01}.

The   model \rf{222}  is  a special case  of the one  for  $\Phi_\a=(\psi, \p)$   with the Lagrangian 
\be 
L= h_{\a\b} \P_\a \Box \P_\b   + m_{\a\b} \P_\a \P_\b   +  V( \del \P) \ , \la{322}
\ee 
where  the constant matrix  $h_{\a\b}$    is not  positive  definite 
 (has  (1,-1) signature) and  $m_{\a\b}$   is degenerate, so  that the kinetic operator 
$h_{\a\b}\Box  +  m_{\a\b} $ is not diagonalizable. 

 If we add to \rf{222}  an extra term $\mu  \p \Box \p $  
 (so that  integrating out $\psi$ gives $ \p\Box^2 \p + \mu \p \Box \p$)
 we get a model   where the kinetic operator can  be diagonalized as
 \be  \la{3335}
  2  \psi \Box \p   - \psi^2   + \m\, \p \Box \p  + V( \del \p)  =
  \m\, \vp \Box \vp - \m^{-1} \psi \Box \psi - \psi^2   + V(\del  \vp - \m^{-1} \del \psi)\, , \ \ \ \ \ \ 
  \vp=\p + \m^{-1} \psi \ . 
  \ee
It thus describes  a diagonal  combination of a  physical massless scalar  and a massive ghost that are mixed   in
the interaction  potential.  
  Focussing just  on the  massless sector of the S-matrix  here   is not justified  a priori. 
The   ``non-diagonal"    model  \rf{222} should not,  however,   be seen as a  limit $\mu\to 0$ of the ``diagonal" model \rf{3335}
as this limit  is singular.\foot{
One  may  rescale the fields $\vp$  and $\psi$   by $\mu^{- 1/2}$  and $\mu^{ 1/2}$  respectively
getting $L'=  \vp' \Box \vp' -  \psi' \Box \psi '-\mu^2  \psi'^2   +  g \mu^{-2} [\del ( \vp' - \psi')]^4$
 and then take the limit $\mu\to 0$  and $g \to 0$  keeping $g'= g \mu^{-2}$   fixed  (we thank J. Donoghue for this remark). However, 
  this theory  can not be viewed as  a smooth limit of the original $(\Box \phi )^2 + g(\del \phi)^4$ 
 theory where $g$ was  non-zero.
 Indeed, setting $\mu=0$ we get $L'=  \vp' \Box \vp' -  \psi' \Box \psi '  +  g' [\del ( \vp' - \psi')]^4
 =  u \Box v  + g' (\del u)^4, \ \   u= \vp' - \psi', \ v= \vp' + \psi'$.
 The same  model is found directly from \rf{222}  by taking the scaling limit $\p= \mu^{-1/2} \p', \ \psi= \mu^{1/2} \psi', \
 g= \mu^2 g', \ \mu\to 0$ that removes the $\psi^2$ term:  
 $L'= 2  \psi' \del^2  \p'      + g'  (  \del \p')^2$.  Introducing the source terms $j_{\p'} \psi' + j_{\psi'} \p'$ and observing that  integrating out 
 $\psi'$   gives the delta-function of $2\del^2  \p'   -  j_{\p'} $ one finds that the generating function 
 contains only the quartic interaction term, i.e. all  quantum corrections in this theory  vanish. 
 This is a reflection of the fact that  in the original theory  $g \to 0$ in this limit.
 } 
Thus the  non-unitarity 
 of the ``dipole ghost"   model  should be  analysed  separately.\foot{This  means, in particular, 
 that  discussions \ci{rr}  
 of the issue of  (non)unitarity  of  $R+ R^2$ type gravity  and similar models  with  massive ghosts 
  may   not  directly   apply to  the present case.}


   Let us first address the question  of which  are the natural  asymptotic  states
in the model   \rf{222}. 
The equations of motion  following from \rf{222}  read
\be \la{111}
\Box \p -\psi =0 \ , \ \ \ \ \ \ \   \Box \psi   - 2 g \del^m \big( \del_m \p (\del \p)^2\big) =0 \ . 
\ee 
To compute,  e.g., tree-level S-matrix we  may solve the classical equations of motion with some 
``in"   boundary condition,   evaluate the action on this solution   and expand in powers on ``in" fields. 
Starting  with the free-theory solution  with non-zero 
 $\psi_0=\psi_{\rm in} =  \int d^4 p\  \delta(p^2)\  \td \psi_{{\rm in}}(p) \ e^{ip\cdot x}$,  $ \Box \psi_{\rm in}=0$,  we get for $\p$
 \ba
& \p_0 = \int d^4 p\  \delta(p^2) \   N_n (p)\  x^n\  e^{ip\cdot  x}      \ , \ \ \ \  \Box^2 \p_0 =0\ , \ \ \ \ 
\del_n \p_0 = \int d^4 p\  \delta(p^2) \   \big[ N_n (p)   + i p_n   N_m(p)\,  x^m\big]   \  e^{ip\cdot  x}  \ , 
 \no    \\
& \Box \p_0 =  2 i \int d^4 p\  \delta(p^2) \  N^n (p) \, p_n \ e^{ip\cdot  x}  =\psi_{\rm in}     \ , \ \ \  \ \ \ \ \ 2 i N^n (p)  p_n 
 \equiv  \td \psi_{{\rm in}}(p)\not=0 \ .  \la{433}
 \end{align}
 Since $\del_n \p_0$    contains a growing part,  the resulting tree-level scattering amplitude with $\psi_{\rm in}$-legs 
  determined  from the  action $g \int d^4x ( \del\p_0)^4$  (cf. \rf{222}) will 
  not  be well defined (in particular, there will be no  usual momentum conservation
   \ci{Adamo:2018srx}   like in  scattering in  some   
  non-translationally invariant background).  

We should  thus   assume   that  $\psi$ should  not have a non-trivial free-theory part, i.e. that
 the asymptotic  field configuration   should be    the  purely-oscillating solution for $\p$:  
\be \psi_0=0\ , \qquad \quad   \p_0=\pn , \qquad  \Box \pn =0  \ , \qquad \pn(x) = \int d^4 p\  \delta(p^2) 
\  \td \p_{\rm in} (p) \  e^{ip x} \ .  \la{344} \ee
Then the resulting   solution  of \rf{111}  will be
\be 
\p = \pn  -  \Box^{-1} \psi, \  \qquad \ \ \   \psi= 2 g \Box^{-1} \del_m \big( \del_m \pn   (\del \pn )^2\big) +   \OO((\del \pn )^5) \ . 
\ee
Plugging this back  into  \rf{222}    will give
\be 
L= g  (  \del  \pn )^4  +  \OO((\del \pn) ^6) \ ,  \ee
which is   of course   the same as what  we  get  by   starting  directly with the $\Box^2$ 
theory  \rf{01} with  the oscillating  mode ($ \Box \pn =0$)   as an asymptotic  state. 

The corresponding   massless  4-particle  scattering amplitude is  then the same
 as in the $\p\Box\p$ theory \rf{02}
 \ba 
  &A^{ (\rm tree)}_4  \sim   g   (s^2 + t^2 + u^2)\la {411}\ ,  \\
 s=- (p_1 + p_2)^2, \ \ \ \  t= - (p_1+p_4)^2, & \ \ \ \ \ u= - (p_1 + p_3)^2 , \ \ \ \ \ \ \   p_i^2 =0 , \ \ \ \ \   \sum_i p_i=0 \ . \no
  \end{align}
   Since 
    the  ``mixed" $(\p,\psi)$ amplitudes are  also IR-singular  one may hope 
    that the ``restricted"  massless S-matrix (the one in the  oscillating $\p$-sector only)  
   constructed by the standard rules  starting with \rf{222}   may  somehow be  consistent  on its own.\foot{Note that in terms  of the original 4-derivative formulation \rf{01}   this implies, in particular,  that the external $\Box^{-2}$  propagators are to be cut off.
 If we define  the  tree-level S-matrix as the classical action evaluated on classical solution with 
``in"   asymptotic conditions  \cite{Arefeva:1974jv}   
we  set  $\Box^k \p + V'( \p)  = j = \Box^k \p_{\rm in}, \ \    \p= \p_{\rm in} + \Box^{-k} V'(p_{\rm in})+ ... $  (where $k=1$ or 2) and then plug 
this into the action. That will give  just $\sim V(\p_{\rm in})$  so cutting off the  external legs   is realized  automatically
   if the source is for $\p$ itself.}
   


  \iffa 
   and that oscillating part of $\p$ is  to be used as an asymptotic state.} 
  Assuming  the 
   ``bad"  $\psi$-states  can not   be produced in scattering of $\p$-fields\foot{This is somewhat 
   analogous to the discussion in   \cite{Hawking:2001yt} that advocated  that starting 
   with Euclidean path integral to define theory   with massive ghost  one should not 
   have ghosts  produced in scattering of physical states. Note however, we do not have 
   massive ghost states   in the  present ``dipole ghost"  model.}
   \fi

\subsection{One-loop scattering amplitude}

 The    procedure  of defining massless  S-matrix in  the  $\p$-sector 
 can be extended to loop  level. 
To illustrate  that the loop scattering amplitudes for the massless $\p$-particles  in the theory \rf{01} with  derivative interactions
do not have IR   divergences at small virtual   momenta 
(despite the fact that    internal   propagators   are $1/q^4$)
 here  we  will  explicitly compute the  one-loop 
4-point scattering amplitude. 

The  one-loop   amplitude   can be found from  \rf{44} 
  by  expanding to th 4-th order in the background field $\vp=\vpn$ with  $\Box \vpn=0$
  so that external momenta  are subject to $p^2_i=0$, i.e.
  \be 
  \Gamma_1  \to \int d^4 x   \int  { d^4 p_1 \ov (2\pi)^4} ...
  { d^4 p_4 \ov (2\pi)^4} \ 
e^{i x \cdot \sum_i p_i } \,  A_4 (p_1, ...,  p_4)\,  \td \vp_{\rm in} (p_1)\ ...\  \td \vp_{\rm in} (p_4)\la{x1} \ .
  \ee
Expanding $\Gamma_1$ in \rf{44} to the  relevant $V^2$ order   we get\foot{We shall first 
use the Euclidean  formulation  (treating  the external momenta  as complex satisfying $p^2_i=0$) 
and rotate to the Minkowski signature  with real $p_i$  only  after computing the loop integrals. 
 Note that here  there is no one-loop 2-point function 
correction  as the  corresponding diagram  is  quadratically divergent and thus can be set to zero  in dimensional regularization that we shall use below.}
\begin{figure}[t]
      \centering
      \includegraphics[scale=0.8]{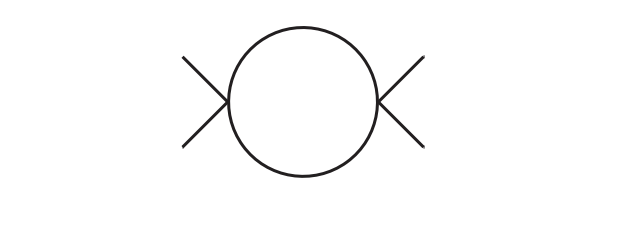}
      \caption[a]{\small  One-loop scattering amplitude. 
      }
\end{figure}
\begin{align} \la{41}
&\Gamma^{(4)}_1 = - { \tf{1}{4}{ \Tr} \Big[ \Box^{-2} \del_m  (V^{mn} \del_n)   \Box^{-2} \del_k( V^{kr} \del_r)\Big] }
\to  
\\
 \la{42}
& A_4 \sim   \VV^{mn} (p_1,p_2 ) \, \VV^{kr} ( p_3,p_4)
 \int { d^4 q\ov (2\pi)^4}    { (p+q)_m     q_n     (p+q)_k    q_r  \ov (p+q)^4\ q^4 }  \ ( 2 \pi)^4 \delta^{(4)} (\sum_i p_i )\ , 
\end{align}
where $p\equiv  p_1 + p_2=-(p_3 + p_4)$  and $\VV_{mn} $ are  given by the  \rf{3}  with the  $\vpn$ factors   stripped off, 
  i.e.  
\be \la{666}
\VV_{mn} (p_1, p_2) \to   2 g (   \delta_{mn}  p_1 \cdot p_2    +  p_{1m} p_{2n} +  p_{2n} p_{1m} ) \ , \ \ \ \ \ 
p^2_{1}= p^2_2=0 \ . 
\ee
The integral in \rf{42} is thus  IR finite  and   contains the same  log  UV divergence   as in \rf{33}.\foot{For large $q$ we get 
$ \int { d^4 q\ov (2\pi)^4}    {1 \ov (q+p)^4\ q^4 }  q_i     q_j     q_k    q_r  \to 
( \d_{ij} \d_{kr} + \d_{ij} \d_{kr}  + \d_{ij} \d_{kr} )  \int { d^4 q\ov (2\pi)^4}    {1 \ov  q^4 }  + ...
$
and that leads to the  UV divergence   proportional to the  combination $2  V^{mn} V_{mn} + (V_{k}^{k})^2  $  as in \rf{311}. }

Renormalizing $g$ in the sum  of the tree \rf{411} and  the one-loop \rf{42}  amplitudes 
we will find that the  latter    will  have the following structure   ($s+t+u=0$)
\be 
A_4  \sim     g^2 \Big[  (s^2+t^2 + u^2)  \log (-{ s\ov \mu^2})   +    s^2 F( { s\ov t} ) \Big] \la{4545}\ ,
 \ee
 where $\mu$ is a renormalization scale. 
As  the classical  theory \rf{01}  is   scale-invariant  ($g$ is dimensionless) 
  the one-loop amplitude     scales  as $s^2$  like the tree-level one 
 in \rf{411}. 
 Our aim  is to compute \rf{4545} explicitly   and then   see   how it  
 indicates  a breakdown of  perturbative unitarity in this theory. 
 

From \rf{666}    we get  (using $p^2_1=p^2_2=0$  and ignoring the $2g$ factor) 
\begin{align}
\VV^{mn} (p_1,p_2) (p_1 + p_2 +q)_m     q_n\ \ \   \to \ \ \  &
 2 p_1 \cdot q\   p_2 \cdot q  + p_1 \cdot p_2 (p_1 + p_2 + q)^2   -  2 (p_1 \cdot p_2)^2 \ .  \la{46}
\end{align}
Using  \rf{42},\rf{46} and the notation
\be\la{644}\te p= p_1 + p_2 = - p_3 - p_4 \ , \ \ \ \qquad   p^2 =  2 p_1 \cdot p_2 = 2 p_3 \cdot p_4 \ ,   
  \ee
 we get  
   the following  expression  for the momentum-dependent   factor $X_4$  in the amplitude 
\ba \la{466}
&A_4 \sim g^2 X_4 \ , \qquad \qquad X_4=  \int { d^4 q\ov (2\pi)^4}    { K(p_1,p_2, q)\ K(p_3,p_4, q)  \ov   (p +q)^4\ q^4 } \ , \\
& K(p_1,p_2, q) \equiv  
   2 p_1 \cdot q\   p_2 \cdot q  +  \ha p^2   (p + q)^2   -  \ha  p^4 
   \ . \la{4666}
\end{align}
It is straightforward    to  compute the integral in \rf{466}  in Euclidean signature   using dimensional regularization.
After evaluating the momentum contractions in \rf{466}  we  find
\ba \la{421}
X_4= &
 4   (p_1\cdot p_2)^4 \ I(2,2)  - 4  (p_1\cdot p_2)^3 \ I(1,2)  +  (p_1\cdot p_2)^2 \ I(0,2)\no \\ 
 &   - 8  (p_1\cdot p_2)^2 p_1^k p_2^l\ I_{kl} (2,2)   + 4  (p_1\cdot p_2) p_1^k p_2^l\ I_{kl} (1,2) 
   + 4 p^i_1 p^j_2 p^k_3 p^l_4\ I_{ijkl} (2,2) 
   \ , 
\end{align}
where the basic integrals $I(n,m), \, I_{kl}(n,m),\, I_{ij kl}(n,m)$  
 are defined in Appendix. 
Using $p_1 \cdot p= p_1 \cdot p_2=-\ha s$,  $p^2= 2 p_1 \cdot p_2 = - s $, $t= -2p_1 \cdot p_4, \ u= -2p_1\cdot p_3$ we get 
\ba
& - 8  (p_1\cdot p_2)^2  p_1^k p_2^l\ I_{kl} (2,2)+ 4  (p_1\cdot p_2) p_1^k p_2^l\ I_{kl} (1,2) \no \\ &\qquad \qquad = 
- \ha  s^4  \big[A_1(2,2)  +   A_2(2,2) \big] \ I(2,2) -    \ha  s^3  \big[A_1(1,2)  +   A_2(1,2) \big] \ I(1,2)\no \ ,   \\
&
4 p^i_1 p^j_2 p^k_3 p^l_4\  I_{ijkl} (2,2) 
= \tfrac{1}{4 }  \Big[ s^4 B_1  (2,2)  + 4s^4  
B_2(2,2)  + s^2  ( s^2 + t^2 + u^2     ) B_3(2,2) 
\Big] \ I(2,2) \la{4200}\ , 
\end{align}
where the expressions  for  $A_k (n,m)$ and $ B_k (n,m)$ are  given in Appendix. 
 Then  \rf{421} becomes 
\ba
X_4= &
  \tfrac{1}{4 }  s^4  I(2,2)  + \ha  s^3 I(1,2)  +  \tfrac{1}{4 }  s^2  I(0,2)  \no \\ &  - \ha  s^4  [A_1(2,2)  +   A_2(2,2) ] I(2,2) -\ha  s^3  [A_1(1,2)  +   A_2(1,2) ] I(1,2)
 \no \\ &  +  \tfrac{1}{4} \Big[ s^4 B_1  (2,2)
+  4 s^4
B_2(2,2) 
+ s^2  ( s^2 + t^2 + u^2     ) B_3(2,2) 
\Big] I(2,2)\la{4211} 
   \ . 
\end{align}
The evaluation   of  the limit $w-2\equiv  {d-4\ov 2} \to 0$   gives  for this $s$-channel  amplitude\foot{
Note  that 
 $4 { t^3 + u^3 \ov s} $ can be written  using   $t+u=-s$  as  $ -4(   t^2 - t u + u^2)   $. 
It is useful to recall that 
for  the  di-gamma   function $\psi_0(z)=  {\rm PolyGamma}[0, z]$ one has 
$\psi_0(n + \ha) = - \g_E - 2 \log 2 + \sum^n_{k=1} {1\ov 2k-1}$.
}
\ba 
 X_4= - \tfrac{1}{96(2\pi)^2} \Big[   ( 13s^2+t^2+u^2) \big[   \tfrac{1}{w-2} + \log (-s)  + \g_E -   \log (4 \pi)   \big]
 -   \tfrac{1}{3}    (32   s^2   +5  t^2 +5  u^2) 
   \Big] \ .   
\la{423}
\end{align}
Symmetrizing in $s,t,u$ we  get  for the full one-loop amplitude 
\ba \la{424}
A_4 \sim  g^2  X_{4}^{(\rm sym)}=  - \tfrac{g^2}{96(2\pi)^2}  \Big[& 15   ( s^2+t^2+u^2)  \big[   \tfrac{1}{w-2} + \log (-s)  + \g_E -   \log (4 \pi)    - \tfrac{14}{15} \big]\no\\
  & + ( 13t^2+u^2+s^2)\log {t\ov s}   +  ( 13u^2+s^2+t^2)\log {u\ov s}  \Big] \ . 
\end{align} 
Here the coefficient of the UV divergent term  agrees  with \rf{33},\rf{4} 
 (extra  factor  of $1\ov 2$   is related to $ \frac{1}{w-2} = {2\ov d-4} \to 2 \log \Lambda$). 
 The divergence is absorbed into the renormalization of $g$ in tree-level amplitude \rf{411}.  
The coefficient of the ``tree-level"  combination $s^2+t^2 +u^2$  is then scheme-dependent.
 After the renormalization we get  the first term in \rf{424}  in  the form $\sim  (s^2 + t^2 + u^2) \log (-{s\ov \mu^2})$
as anticipated  in \rf{4545}.

 Note  that  in   the  familiar  massless $\p\Box\p  + g \p^4$ theory the  corresponding 
  result for the  $s$-channel one-loop   amplitude in Fig. 1  is  proportional to (cf. \rf{413}) 
\be\la{426}  I(1,1)\big|_{w\to 2} = -\tfrac{1}{16 \pi ^2}\big[  \tfrac{1}{w-2} + \log (- s )   + \g_E - \log (4 \pi) -2  \big] \ .  \ee
Let us  also  compare the  expression \rf{424}  to the one-loop amplitude   in the perturbatively  unitary  
 but 
non-renormalizable   analog \rf{02} of  the model \rf{01}.
Here  the one-loop  massless amplitude is  given by  the same expression as in \rf{466}   but with the standard 
 $1/q^2$ propagators:
 \be \la{4660}
A_4 \sim \bg^2 X_4 \ , \qquad  \ \ \  X_4=
\int { d^4 q\ov (2\pi)^4}  
{  K(p_1,p_2, q) \, K(p_3,p_4, q) \ov   (p +q)^2\ q^2 }\ . 
\ee
Using \rf{413}--\rf{418}  instead of \rf{4211} the  counterpart of \rf{4211}  then reads\foot{Note that  
 here the arguments of the  integrals $I(n,m)$ and $A_i(n,m), \ B_i(n,m)$  are   shifted by 1 compared to \rf{4211}.}
\ba
X_4= &
 \tfrac{1}{4}  s^4 \ I(1,1)  + \ha  s^3\ I(0,1)  +  \tfrac{1}{4}  s^2\ I(-1,1)  \no \\ &  - \ha  s^4  \big[A_1(1,1)  +   A_2(1,1) \big] \ I(1,1) -
 \ha  s^3  \big[A_1(0,1)  +   A_2(0,1) \big]\  I(0,1)
 \no \\ &  +   \tfrac{1}{4}  \Big[ s^4 B_1  (1,1)
+ 4 s^4   
B_2(1,1) 
+ s^2  ( s^2 + t^2 + u^2     ) B_3(1,1)  \Big]\  I(1,1)\la{42110} 
   \ . 
\end{align}
Taking the  limit $w\to 2$ 
gives   for the  $s$-channel  
and symmetrized  in $s,t,u$  amplitudes (cf.  \rf{423})  
\ba
X_4= - \tfrac{1}{ 960(2 \pi)^2  }\Big(  &   s^2 \big( 41 s^2  + t^2 +u^2  \big)
 \big[ \tfrac{ 1}{ w-2} +  \log (-s)   + \gamma_E - \log (4  \pi) \big]\no \\ &
   -  \tfrac{1}{15} s^2 \big[ 1301 s^2  - 23(t^2 + u^2)  \big] \Big) \, ,
  \la{348} \\
 X_{4}^{( \rm sym)}= - \tfrac{1}{ 960(2 \pi)^2  }\Big( &     \big[ 41 (s^4 + t^4 + u^4) + 2( s^2t^2 +s^2u^2+ t^2 u^2)  \big]
 \big[ \tfrac{ 1}{ w-2} +  \log (- s)   + \gamma_E - \log (4  \pi) \big]  \no \\
 &   + t^2  \big( 41 t^2  + u^2 +s^2  \big)   \log {t\ov s}    +
 u^2\big( 41 u^4  + s^2 +t^2  \big)   \log {u\ov s}  \no \\
 & 
  - \tfrac{1}{15} \big[ 1301 (s^4 +t^4 + u^4) - 46 (s^2t^2 + u^2t^2 + u^2 s^2)  \big] \Big) \ . 
  \la{3480}
\end{align}
The expression \rf{3480}  can be simplified 
 using     that 
$  s^4 + t^4 + u^4= 2( s^2t^2 +s^2u^2+ t^2 u^2)=\ha (s^2 +t^2 + u^2)^2 $ 
 and   finally we get (cf. \rf{424}) 
\ba
A_4\sim \bg^2X_{4}^{( \rm sym)}= - \tfrac{\bg^2}{960 (2\pi)^2  }\Big[ &      21 (s^2 + t^2 + u^2)^2 
 \big[ \tfrac{ 1}{ w-2} +  \log (- s)   + \gamma_E - \log (4  \pi) - \tfrac{213}{5} \big]  \no \\
 &  + t^2  \big( 41 t^2  + u^2 +s^2  \big)   \log {t\ov s}    +
 u^2\big( 41 u^4  + s^2 +t^2  \big)   \log {u\ov s} 
  \Big]\ .  
  \la{3488}
\end{align}
\iffa
\ba
X_4=  \tfrac{1}{ 480(2 \pi)^2  }\Big(  & -   \big[ 44 s^4  + 10 s^2(t^2 +u^2)  \big]
 \big[ \tfrac{ 1}{ w-2} +  \log (-s)   + \gamma_E - \log (4  \pi) \big]\no \\ &  + \tfrac{1}{15} \big[ 1424 s^4 + 405s^2(t^2 + u^2)  \big] \Big) \ ,
  \la{348} \\
 X_{4}^{( \rm sym)}=  \tfrac{1}{ 480 (2 \pi)^2  }\Big( &  -   \big[ 11 (s^4 + t^4 + u^4) + 5( s^2t^2 +s^2u^2+ t^2 u^2)  \big]
 \big[ \tfrac{ 1}{ w-2} +  \log (- s)   + \gamma_E - \log (4  \pi) \big]  \no \\
 &   + 2 \big[ 22 t^4  + 5 t^2(u^2 +s^2)  \big]   \log {t\ov s}    + 
 2\big[ 22 u^4  + 5 u^2(s^2 +t^2)  \big]   \log {u\ov s}  \no \\
 & 
  + \tfrac{2}{15} \big[ 712 (s^4 +t^4 + u^4) + 405 (s^2t^2 + u^2t^2 + u^2 s^2)  \big] \Big) \ . 
  \la{3480}
\end{align}
The expression \rf{3480}  can be simplified 
 using     that 
$ \ha (s^2 +t^2 + u^2)^2 =  s^4 + t^4 + u^4$= $2( s^2t^2 +s^2u^2+ t^2 u^2)=2 (t^2+ tu + u^2)^2
$
 and   finally we get (cf. \rf{424}) 
\ba
A_4\sim \bg^2X_{4}^{( \rm sym)}=  \tfrac{\bg^2}{480 (2\pi)^2  }\Big( &  -    27 (s^2 + t^2 + u^2)^2 
 \big[ \tfrac{ 1}{ w-2} +  \log (- s)   + \gamma_E - \log (4  \pi) - \tfrac{1609}{30} \big]  \no \\
 &   + 2 \big[ 22 t^4  + 5 t^2(u^2 +s^2)  \big]   \log {t\ov s}    + 
 2\big[ 22 u^4  + 5 u^2(s^2 +t^2)  \big]   \log {u\ov s} \Big)\ .  
  \la{3488}
\end{align}
\fi
The  UV  divergence here   scales as  8-th  power of momentum 
  so that   to cancel it   one   needs    a  new    counterterm   $\sim(\del \del \p)^4$ to be added to the  action \rf{02}.

The imaginary part  of  the amplitude \rf{3488}   comes from $\log(- s)  \to \log |s| + i \pi$ and thus 
has the same structure as the square $(s^2 + t^2 + u^2)^2$  of the  tree-level   amplitude \rf{411}.
This is  
 in agreement with the generalized optical theorem 
(see,  e.g., \cite{Peskin:1995ev}):  Im part of one-loop amplitude 
  should   be found   by  multiplying  two tree-level amplitudes 
and summing over   intermediate on-shell states with the standard phase space measure $\sim \int d^4 q\, \delta(q^2)$. 
This is  of course  the  same  conclusion that one reaches  in the standard $\phi \Box \phi + g \phi^4$ theory 
(cf. \rf{426}). 

This  suggests   that  the optical theorem directly  related to unitarity  should  
 fail in the 4-derivative model \rf{01}  where the imaginary part of \rf{424} is proportional to the tree level amplitude \rf{411}  itself
 rather than its  square  (this fact is   directly related to its renormalizablity).
We shall  elaborate on this  unitarity issue   in the next subsection. 



\subsection{(Non)unitarity}

The  unitarity of the standard $\p \Box \p + g \p^4$   theory  manifests  itself in the relation 
between the imaginary part of the corresponding one-loop amplitude \rf{426} coming 
 from\foot{After  continuation to Minkowski   $(-+++)$ signature 
 $s=- (p_1 + p_2)^2 $ is the c.o.m.   energy that  should be  positive.}
  \be   \log(- s)  \to \log |s| + i \pi \ ,  \la{xxx} \ee 
 and the square of  the (constant) tree-level   amplitude, 
 i.e.  in the generalized optical theorem. 

The  same   applies  also to the  $\p \Box \p + \bg (\del \p)^4$ theory \rf{02}. 
While the  $s^2$ growth of the corresponding 
tree level  amplitude \rf{411} is indicating a conflict with unitarity  at large  energies, 
this theory is  unitary  in perturbation theory at  small enough energies  $s < M^2$  \cite{Weinberg:1995mt,Adams:2006sv}. 
An indication of this  is  that  the imaginary   part of the one-loop  scattering amplitude \rf{3488}
is proportional to the square of the tree-level amplitude  \rf{411} (see also below). 

At the same time, the imaginary part of the one-loop 
scattering   amplitude  \rf{424}  in the scale-invariant $\p \Box^2  \p + g (\del \p)^4$ theory with dimensionless  coupling 
is proportional to the first power of the tree-level  amplitude \rf{411}  suggesting  violation of  the 
 generalized optical theorem. 

\iffa More precisely,  the imaginary   part of the 
one-loop amplitude  should be related 
 to  the integral of the product of two  tree-level amplitudes over the phase space 
of on-shell   intermediate states.
\fi  
Let us  recall  the   general 
argument   relating the unitarity of  the S-matrix to the   generalized     optical   theorem \cite{Peskin:1995ev}
(see also \cite{Buoninfante:2022krn}).
 Given    ${\rm S}= 1 + iT$,  taking  the  matrix element of the unitarity relation $- i( T-T^\dagger) = T^\dagger  T$ 
    between the  two 2-particle states of massless particles  $ \langle1,2|T|3,4\rangle=A_4(p_1, ..., p_4)\   \delta^{(4)}(\sum_i p_i ) $
gives
\ba &-  i \big[ A_4(p_1,p_2 \to p_3, p_4) - A^*_4(p_3,p_4 \to p_1, p_2) \big]  \la{332} \\
&= \sum_n \prod^n_{i=1} \int {d^4 q_i \ov (2 \pi)^4}\ \delta(q^2_i) \   A_4(p_1,p_2 \to q_1,.., q_n) \ A^*_4(p_3,p_4 \to q_1,.., q_n) \
( 2 \pi)^4 \delta^{(4)}(p_1 + p_2 - \sum^n_{i=1} q_i)\no \ . 
\end{align}
In the  present   case of the 
 one-loop diagram in Fig.1   with $n=2$ internal propagators  
 the r.h.s. of \rf{332}  should come from ``cutting"  the propagators in  
$ \int d^4 q  {1\ov  q^2 (q+p)^2}= \int d^4 q_1 \int d^4 q_2  {1\ov  q_1^2   q_2^2}\delta^{(4)} ( p- q_1 -q_2 )  $
 using 
\be 2\,  \Im {1\ov q^2 - i \eps} =     {2 i \eps \ov (q^2 + \eps^2)^2} \to   2 \pi i \delta(q^2)\ . \la{777} \ee
 In the  standard $\p^4$   theory the  tree-level  amplitude  is just a  momentum-independent constant  and one   can 
   verify  that 
\ba \la{333}
&2\,  \Im A^{(one-loop)}_4(p_1,p_2,p_3,p_4)  = 
   \int d \Pi(q_1) \,  \int d \Pi(q_2) \no \\
& \times A_4^{ (\rm tree)}( p_1,p_2, q_1, q_2) \  A_4^{ (\rm tree)}( q_1,q_2, p_3, p_4) \  ( 2 \pi)^4 \delta^{(4)}(p_1 + p_2 - q_1 - q_2)\ , 
\end{align}
where  $ \int {d^4 q \ov (2 \pi)^4}2\pi \delta(q^2) \theta ( q^0)  \equiv  \int d \Pi(q) =
 \int {d^3 \vec q \ov (2 \pi)^3 \ 2| \vec q|}$.
 Indeed,  from\foot{Note  that  in continuing from the Euclidean  to Minkowski  signature
  we get  an extra $i$ factor  from 
 $\int d q^0 \to i \int d q^0$.  In general, 
  one has ${1\ov q^2 - i \eps} = \PP {1\ov q^2}  + i \pi \delta(q^2) $   so that the imaginary part of the one-loop  diagram  comes from both $ - \pi^2 \delta(q^2)\, \delta((q+p)^2) $  
   and the  principal value  product $ \PP {1\ov q^2}\  \PP {1\ov (q+p)^2}$   but they give similar 
  contributions. We thank R. Roiban for  a clarifying discussion of  this point.}
 $\int {d^4q \ov  q^2 (q+p)^2} \to \int d^4 q\  \delta(q^2)\ \delta((q+p)^2) $ 
 using  the c.o.m. frame where $p=p_1 + p_2 = (p^0, 0,0,0)=\sqrt{s}$  we get for the imaginary part\foot{The positivity of $q^0$ 
 follows here  from positivity of $p^0$.}
 \ba \int   {d^4 q }{1 \ov   (q^2 - i \eps)  ((q+p)^2 - i \eps)} \to & \int d^4 q\,   \delta(q^2)\,  \delta((q+p)^2)\no \\
  \to 
& \ 4\pi  \int dq^0\ \int d| \vec q| \ | \vec q|^2\  {  \delta ( q^0 - | \vec q|)\ov 2 | \vec q|}\  \delta ( ({p^0}{})^2 - 2 p^0 q^0) 
\to  \pi\ , \la{99}\end{align} 
  and this  should match   the constant  Im part of \rf{426}.

The same   argument  applies  in the case of the $\p \Box \p + \bg (\del \p)^4$   theory:  rotating  the one-loop amplitude \rf{4660} to the Minkowski signature  with $i \eps$ prescription for the propagators 
and using the cutting rule \rf{777} we get in the c.o.m.   frame
 \ba  
&  \int d^4 q\,   \delta(q^2)\,  \delta((q+p)^2) \   K(p_1, p_2, q) \ K(p_3, p_4, q) 
  \no \\
  \to 
& \ \four \int  dq^0\  \int d^3 \vec q \   {  \delta ( q^0 - | \vec q|) }\ 
 \delta (q^0 - \ha p^0) \  K(p_1, p_2, q) \ K(p_3, p_4, q) \ .  
\la{909}\end{align} 
In the  c.o.m. frame  with  $p_1= ( \ha p^0, \vec p_1) , \ 
p_2= ( \ha p^0, - \vec p_1) , \ p_3= ( \ha p^0, \vec p_1) , \ 
p_4= ( \ha p^0,  -\vec p_1) $
corresponding to forward scattering ($t=0$, $s=-u$) 
this  reduces to the imaginary part of the one-loop  amplitude 
in \rf{348}\foot{Integrating $[K(p_1, p_2, q) ]^2$ where $K(p_1, p_2, q) = 
  2 p_1 \cdot q\   p_2 \cdot q  - \ha p^2   (p + q)^2   -  \ha  p^4$    (cf. \rf{4666})  
  over $\vec q$ one is to use that $\int d^3 \vec q \,   q_i q_j ( ...) = {1\ov 3} \int d^3 q   |\vec q|^2  \delta_{ij}  ( ...)$ 
  and similar relation for $\int d^3 \vec q\,    q_i q_j q_k q_l ( ...)$. }
      as found according to \rf{xxx}.   On the  other hand,  eq.\rf{909} has the form of the  phase space integral \rf{333}
of the product of the  corresponding tree-level amplitudes \rf{411}.

In    the $\p \Box^2\p + g (\del \p)^4 $ theory the 
 unitarity   of the massless  S-matrix 
  may     not be expected  a priori 
   as we  consider    only  massless ($\Box \p=0$)    states  on  the external lines  but  have 
 $1/q^4$ instead of $1/q^2$  internal  propagators (implying that effectively there  are 
 more  virtual ``states" than just massless  ones).   
  The argument  leading  to the generalized optical theorem  \rf{333}  uses the insertion of  a complete set of states 
  ($1 = \sum |...\rangle \langle ...|$) between the  $T$-matrix and its conjugate.
  This  would effectively mean summing also over the  ``growing" modes (or $\psi$-states in \rf{222})  
  but the  corresponding tree-level   amplitudes are not well defined.

 To see the problem with checking the  optical theorem 
 more explicitly
  we need  to define  the Minkowski-space analog of the  $\Box^{-2} \sim \log x^2$  or
 its formal Fourier  image  $1/q^4$. 
A  natural starting point is the  2-derivative   formulation \rf{222}: if we formally  replace there 
$\Box \to \Box  + i \eps$ (or $q^2 \to q^2 - i \eps$) that   will  be equivalent, upon integrating out the auxiliary field 
  $\psi$, 
to the following  prescription for the internal propagators 
\be 
{1\ov q^4} \ \ \to \ \  {1\ov (q^2 - i \eps)^2} \ . \la{434}\ee
This prescription  turns out to be  consistent 
with the one  used  in a similar 4-derivative vector  model in   \cite{dEmilio:1982ghe}.
The suggestion there was  that the   Fourier image  of $\Box^{-2} \sim \log  (- \mu^2 x^2 + i \eps )   $ 
should be defined as\foot{The parameter   $a$ here   is  needed  \cite{dEmilio:1982ghe} to define
the  relevant  distribution. This prescription  was  claimed in  \cite{dEmilio:1982ghe}  to be  consistent with causality
and Wick rotation.} 
\be \Box^{-2 } \ \ \  \to \ \ \ 
{1\ov (q^2 - a^2 - i \eps)^2}    + i \pi \log{a^2\ov \mu^2}\,   \delta^{(4)}(q)   \ , \ \ \ \ \ a \to 0
 \ . \la{355} \ee 
In  the  present theory \rf{01}  with only  derivative-dependent interactions  
 the  contact $\delta^{(4)}(q) $ term  in \rf{355}   drops out  of the Feynman diagrams and 
thus  we end up with the same prescription as in \rf{434}. 
 From  \rf{434} we then find the following analog of the standard cut discontinuity relation \rf{777}
\be\la{336} 
  {1\ov (q^2 - i \eps)^{2}}-  {1\ov (q^2 +i  \eps)^{2}}   = 
  {4 i \eps  q^2 \ov [(q^2)^2 +  \eps^2]^{2} }
 \to     - 2 \pi i   \delta'(q^2)  \ , \ \ \ \  \ \ \ \  \delta'(q^2)  \equiv  {\del \ov \del q^2}  \delta(q^2) 
 \   ,\ee
where we used  that  
 $ \delta(x) = {1\ov \pi}   {\eps \ov x^2 + \eps^2}\big|_{\eps\to 0} , \ \   \delta'(x) = -   {1\ov \pi}   {2 \eps  x \ov (x^2 + \eps^2)^2}\big|_{\eps\to 0} $. 
Note   that  \rf{336}  is the  same as the $q^2$    derivative of \rf{777}.
Using \rf{336} in the general expression for   one-loop amplitude \rf{42}  
we   will get  (in the case of forward scattering)   
\be \la{3400}
\VV_{ij}(p)  \VV_{kr}(-p) \int {d^4 q } {d^4 q'}  \delta^{(4)}(  q  + p -  q' )  \  q_j q_r   {\del \ov \del q^2} \delta(q^2)  \ q'_i q'_k  {\del \ov \del q'^2} \ \delta(q'^2)  \ . 
\ee 
\iffa
It  is not immediately 
 clear  how to show that the resulting  imaginary part of Minkowski-space amplitude will be the same as following directly from 
 \rf{424}   and  \rf{xxx}: this requires  using  factors of  derivatives  of delta-functions to remove 4  extra 
 powers of external momenta in the numerator. 
 \fi
 Having the derivatives  of delta-functions  instead of delta-functions as in \rf{332},\rf{777} 
 precludes one from  going 
 through the same  steps as in the standard proof of the optical theorem.\foot{In particular, it  is not clear  which particular modification of the standard phase space measure these $ {\del \ov \del q^2} \delta(q^2)$ factors may be  effectively equivalent to.}


\iffa 
This  we are to compare  to what we may expect from optical theorem -- integrated square of the tree amplitude 
($q=k_1, \ q'=k_2$) 
\be \la{341}
 \int {d^4 k_1 } {d^4 k_2}  \delta^{(4)}(  k_1 + k_2 - p)\ \delta(k_1^2) \delta(k_2^2) 
 \big[ (p_1\cdot p_2)^2  + (p_1 \cdot k_1)^2 + (p_1 \cdot k_2)^2 \big]^2 
 \ee
 Dimensions  do not   match:  extra $qq$ factors in \rf{340} are balanced against derivatives of delta-functions
 but tree amplitude is squared.
 \fi

\renewcommand{\theequation}{4.\arabic{equation}}
 \setcounter{equation}{0}

\section{Concluding remarks}

The four-derivative model  \rf{01} considered in this paper has no mass parameters  and thus  is distinguished  by its classical scale invariance.
It is   different  from the  Pais-Uhlenbeck type  \cite{Pais:1950za,Smilga:2017arl}  models  $L= \phi (\Box - m_1^2) (\Box- m_2^2) \phi + ...$ 
that can be diagonalized into a system of a physical and ghost-like   massive  fields   with the latter carrying negative energy; once interactions are included,  here the production of ghosts 
 is expected, in general,  to lead to a violation of unitarity.

Writing the quadratic  part of  \rf{01}  in the 2-derivative form \rf{222}  demonstrates that the corresponding non-diagonalizable system  has a   non-positive Hamiltonian  density ($H = 2 p_\psi  p_\phi + \psi^2 +... $). 
Whether that leads to a problem  may   depend on a type of    interactions considered   \ci{Smilga:2017arl}. 
That brings in, in particular,  the question  about  the sign of the coupling $g$ in  \rf{01}. 
The Euclidean  path integral is well defined  for $g>0$ (when the action is non-negative); its direct continuation to Minkowski signature  assuming\foot{We
  use  $t_E=- i t_0$   instead of  the conventional $t_E=+ i t_0$ 
to preserve the positive sign of the ``kinetic" term  in $S_M$.}
$t_E=- i  x^0$ 
 gives $e^{-S_E} = e^{i S_M}$  where  $S_M = \int d^4 x\ L, \ \  L= [ ( \del_0^2 - \del_i^2) \p ]^2  
+ g [ (\del_0\p)^2 -(\del_i \p)^2 ]^2$.
Thus in contrast to the  familiar cases of 2-derivative theories  here the signs of  both the   kinetic and the interaction terms 
are  not reversed  (i.e.  the Minkowski-space  Lagrangian is also positive  if $g >0$). 

The  assumption of $g >0$  leads, however, to the presence of unstable  classical solutions. 
Indeed, let  us consider for simplicity just spatially constant   backgrounds $\p=\p(x^0)$   for which 
$L= \ddot \p^2  + g \dot \p^4= \dot v^2 + g v^4, \ \  v=\dot \phi\equiv \del_0 \p.$ For $g>0$ this describes  an  inverted  anharmonic oscillator  with solutions blowing up  in finite time.  To avoid this singular behaviour one is thus  to choose $g<0$. 
Interestingly, the resulting  quantum 4d  theory is   asymptotically free  for $g <0$ (cf. \rf{5})
and is  thus    well defined at short distances.\foot{Note that the question  about the  sign of $g$ is not relevant in 
perturbation theory. Also, the above discussion does not apply to the perturbatively unitary 
effective  theory \rf{02} where $\bar g >0$ is implied by the causality condition as discussed in  the Introduction.}

Our main focus  in this paper  was on demonstrating   that given that the scattering amplitudes 
are well defined  for the massless  oscillating modes only, the  generalized optical   theorem and thus the perturbative 
unitarity that it follows from is violated in the theory \rf{01}  at the  one-loop level  (regardless the sign of $g$). 
One may wonder  if the unitarity issue may  somehow  be resolved   beyond  the perturbation theory
or by adding extra  interactions  that may modify the ${1\ov q^4}$ propagator at quantum   loop level (cf. \ci{rr}). 

One may also   explore a possibility  that  the S-matrix   for the  dimension 0  scalar field 
is not the right observable in this model. For example, one may  conjecture that  these dimension 0 particles   are
always  ``confined", 
i.e.   do not appear as asymptotic states  but may  still 
participate in interactions with  other fields (like  gravity) through virtual loops.  

One may also  attempt to interpret  the   $\Box^2$  theory as a limit of a   ghost-free   non-local theory  (cf. 
\ci{Buoninfante:2016iuf, Koshelev:2021orf}). For example, starting   with $L=\p \Box{ e^{\eps \Box} -1 \ov \eps} \p  + g (\del \p)^4$ 
where $\eps$ has dimension of (length)$^2$  
one may consider  the   limit $\eps\to 0$. 
\section*{Acknowledgments }
We  would like to thank  R. Roiban   for  useful  comments and  suggestions. 
We also acknowledge    discussions with  S. Kuzenko,   R. Metsaev, K.  Mkrtchyan, A. Smilga   and  A. Tokareva.
 We are also grateful to R. Percacci  for pointing out a mistake in eq.\rf{4200}  in  the original version 
 and to J. Donoghue  for raising a number of important issues.
 This   work  was supported by   the STFC grant ST/T000791/1. 
\newpage
\appendix
\section{Basic one-loop integrals}
\setcounter{equation}{0}
\setcounter{footnote}{0}
\def\theequation{A.\arabic{equation}}
\setcounter{equation}{0}
To reduce \rf{466} to basic integrals  in dimensional  regularization (see, e.g.,  \cite{Smirnov:2006ry}) we   may use  the  standard relations   
\ba
&A^{-n} B^{-m} = {\Gamma(m+n)\ov \Gamma(m) \Gamma(n) } \int^1_0  dx \ x^{n-1} (1-x)^{m-1}   \big[ x  A +  (1-x)  B \big]^{-n-m}\ , \\
&
\int  {  d^{2w}q \ov (2 \pi)^{2w} }   \big( q^2   +  2  P\cdot q + M^2\big)^{-\alpha} = 
{\Gamma(\alpha- w)\ov  (4 \pi)^{w}  \Gamma(\alpha) }   (M^2 - P^2)^{-\alpha + w} \ , \qquad   d= 2w  \ , \end{align}
and $ \int^1_0 dx \ x^{a-1}  (1-x)^b  =  { \G(a) \G(b) \ov \G(a+b) } $. 
In our case:   $A=(q+p)^2, \ B= q^2$, $\alpha= n+m $, 
$(1-x)  B + xA =  q^2  + 2 x p\cdot q + x  p^2$,\  $P= x p, \ \ M^2= x  p^2$ 
  so we get 
\be \la{413}
I(n,m) \equiv  \int  {  d^{2w}q \ov (2 \pi)^{2w} }   \  {1\ov  ((q+p)^2)^n (q^2)^m} 
=  { (p^2)^{w - n -m} \ov (4 \pi)^{w} } \  {\G( w-n) \G(w-m) \G(n+m -w) \ov \G(n) \G(m) \G(2w - n - m ) }\ .
\ee
We need   also the following generalizations of \rf{413}
(cf.   \cite{Leibbrandt:1975dj} for $n=m=1$)\foot{These  follow, e.g., 
  by integrating over the Feynman parameter   the expression  for  $\int  {  d^{2w}q \ov (2 \pi)^{2w} }   \big( q^2   +  2  P\cdot q + M^2\big)^{-\alpha}
  q_i.... q_k $.}  
\begin{align} \la{415}
& I_{ij}(n,m) = \int  {  d^{2w}q \ov (2 \pi)^{2w} }    {q_i q_j \ov    ((q+p)^2)^n (q^2)^m }  =  \Big[ A_1(n,m)\  p_i p_j 
 +  \ha A_2(n,m)\  p^2   \d_{ij}\Big]   I(n,m) \ , \\
&\te A_1(n,m) ={ (w-m +1) (w-m) \ov (2w-m-n+1) ( 2w -m -n) }   \ , \qquad \ \ \  
  A_2(n,m)  =    { (w-m ) (w-n) \ov (2w-m-n+1) ( 2w -m -n) (m+n-w-1)}\  ,\no  
\\\la{416}
I_{ijkl}(n,m)  = &\int  {  d^{2w}q \ov (2 \pi)^{2w} }    {q_i q_j q_k q_l \ov  ((q+p)^2)^n (q^2)^m   } =\Big[  B_1(n,m) \ p_i p_j p_k p_l \,  \no  \\
&
 + \ha  B_2(n,m)\  p^2   (\d_{ij} p_k  p_l   +  \d_{ik} p_j  p_l + \d_{il} p_k  p_j  + \d_{jk} p_i  p_l + \d_{jl} p_k  p_i + \d_{kl} p_i  p_j   )\,  \no \\
& + \four   B_3(n,m) \ p^4  (\d_{ij}\d_{kl}  + \d_{ik}\d_{jl} +  \d_{il}\d_{kj}      ) \Big] \, I(n,m)\ ,  \\
& \te B_1(n,m) =  { (w-m +3) (w-m +2 ) (w-m +1) (w-m ) \ov (2w-m-n+3) ( 2w -m -n +2 ) (2w-m-n+1) ( 2w -m -n ) }   \ , \ \ \ \no \\
&\la{418}
\te B_2(n,m)= {  (w-m +2) (w-m +1  )(w-m) (w-n)  \ov  (2w-m-n+3) ( 2w -m -n +2 ) (2w-m-n+1) ( 2w -m -n )    (n+m - w-1) }  
 \ , \ \\
&\te B_3(n,m) =  { (w-n +1) (w-n  ) (w-m +1) (w-m ) \ov  (2w-m-n+3) ( 2w -m -n +2 ) (2w-m-n+1) ( 2w -m -n )    (n+m - w-1) (n+m - w-2)} .
  \no
\end{align}

\newpage


\end{document}

\bibitem{tHooft:1972tcz}
G.~'t Hooft and M.~J.~G.~Veltman,
``Regularization and Renormalization of Gauge Fields,''
Nucl. Phys. B \textbf{44} (1972), 189-213
\bibitem{Barvinsky:1985an}
A.~O.~Barvinsky and G.~A.~Vilkovisky,
``The Generalized Schwinger-Dewitt Technique in Gauge Theories and Quantum Gravity,''
Phys. Rept. \textbf{119} (1985), 1-74
\bibitem{Bogolyubov:1990kw}
N.~N.~Bogolyubov, A.~A.~Logunov, A.~I.~Oksak and I.~T.~Todorov,
``General principles of quantum field theory,''  
Kluwer, Springer, 1990. doi:10.1007/978-94-009-0491-0